\documentclass[twocolumn]{aastex62}

\usepackage{lineno}

\begin{document}
\citestyle{nature}

\title{Observational evidence for bar formation in disk galaxies via cluster-cluster interaction}

\correspondingauthor{Myungshin Im}
\email{yymx2@astro.snu.ac.kr, mim@astro.snu.ac.kr}

\author[0000-0003-0134-8968]{Yongmin Yoon}
\affiliation{Center for the Exploration of the Origin of the Universe (CEOU),
Astronomy Program, Department of Physics and Astronomy, Seoul National University, 1 Gwanak-ro, Gwanak-gu, Seoul, 151-742, Republic of Korea}

\author[0000-0002-8537-6714]{Myungshin Im}
\affiliation{Center for the Exploration of the Origin of the Universe (CEOU),
Astronomy Program, Department of Physics and Astronomy, Seoul National University, 1 Gwanak-ro, Gwanak-gu, Seoul, 151-742, Republic of Korea}

\author[0000-0001-9501-1252]{Gwang-Ho Lee}
\affiliation{Steward Observatory, University of Arizona, 933 North Cherry Avenue, Tucson, AZ 85721, USA}
\affiliation{Korea Astronomy and Space Science Institute, Daejeon 305-348, Republic of Korea}
\affiliation{KASI-Arizona Fellow}

\author{Seong-Kook Lee}
\affiliation{Center for the Exploration of the Origin of the Universe (CEOU),
Astronomy Program, Department of Physics and Astronomy, Seoul National University, 1 Gwanak-ro, Gwanak-gu, Seoul, 151-742, Republic of Korea}

\author{Gu Lim}
\affiliation{Center for the Exploration of the Origin of the Universe (CEOU),
Astronomy Program, Department of Physics and Astronomy, Seoul National University, 1 Gwanak-ro, Gwanak-gu, Seoul, 151-742, Republic of Korea}

\section*{}

\textbf{Bars are an elongated structure that extends from the centre of galaxies, and about one-third of disk galaxies are known to possess bars \citep{Jogee2004,Marinova2009,Lee2012}. These bars are thought to form either through a physical process inherent in galaxies \citep{Athanassoula1986,Kwak2017,Zana2018}, or through an external process such as galaxy-galaxy interactions \citep{Miwa1998,Berentzen2004,Lokas2014}. However, there are other plausible mechanisms of bar formation that still need to be observationally tested. Here we present the observational evidence that bars can form via cluster-cluster interaction \citep{Bekki1999}. We examined 105 galaxy clusters at redshift $\mathbf{0.015 < z < 0.060}$ that are selected from the Sloan Digital Sky Survey data, and identified 16 interacting clusters. We find that the barred disk-dominated galaxy fraction is about 1.5 times higher in interacting clusters than in clusters with no clear signs of ongoing interaction (42\% versus 27\%). Our result indicates that bars can form through a large-scale violent phenomenon, and cluster-cluster interaction should be considered an important mechanism of bar formation.}

\begin{figure*}
\includegraphics[scale=0.18,angle=00]{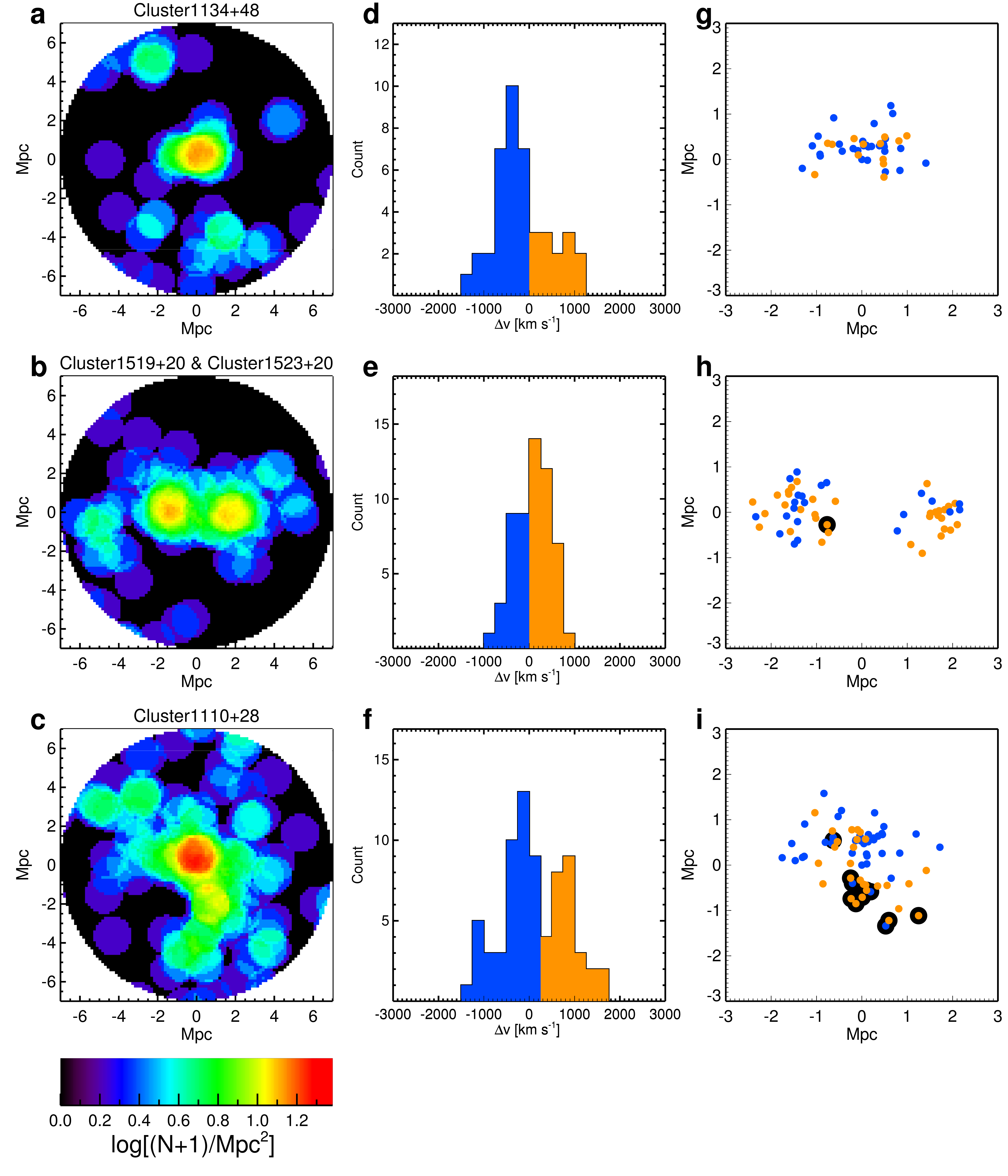}
\centering
\caption{\textbf{Examples of the surface number density of galaxies around clusters, and the velocity and spatial distributions for the cluster member galaxies.} \textbf{a,d,g,} A cluster that is not in a pair and does not have a substructure. \textbf{b,e,h,} A cluster that is in a pair. \textbf{c,f,i,} A cluster that has a substructure. \textbf{a--c,} Maps of the surface number density of galaxies over a rectangular area of 14 Mpc both in right ascension ($x$ axis) and declination ($y$ axis). Each grid size in the $x$ and $y$ directions was set to be 140 kpc. At each point, we measured the surface galaxy number density in an aperture with a radius of 1 Mpc within a rest-frame velocity slice of $\pm2000$ km s$^{-1}$. The colour represents the surface number density (see the color scale in \textbf{c}). \textbf{d--i,} The velocity (\textbf{d--f}) and the spatial (\textbf{g--i}) distributions of cluster member galaxies. We divided the member galaxies into two groups; one with low velocities (blue) and the other with high velocities (orange). The member galaxies with large local deviations of $\delta\ge2.7$ derived by the Dressler-Shectman (DS) test (see Methods) are marked by black circles. For a cluster with a substructure (Cluster1110+28), galaxies of the two groups have different spatial distributions and the member galaxies with large local deviations in velocity are concentrated in a specific area.
\label{fig:exmap}}
\end{figure*}

We used a volume-limited sample of galaxies with $\log(M_\mathrm{star}/M_{\odot}) \ge 10.0$ in the MPA-JHU catalogue, and selected 105 clusters with $M_{200}>7\times10^{13}\,M_{\odot}$ in the redshift range of $0.015<z<0.060$. Here, $M_\mathrm{star}$ is the galaxy stellar mass, $M_{\odot}$ the mass of Sun and $M_{200}$ the cluster halo mass (see Methods). Of these, 16 clusters were found to be in pairs or have substructures (see Methods) and are defined here as interacting clusters. Figure \ref{fig:exmap} shows examples of the surface number density maps, velocities, and spatial distributions for galaxies of clusters in isolation, in pairs, and with substructures.
 
\begin{figure*}
\includegraphics[scale=0.60,angle=00]{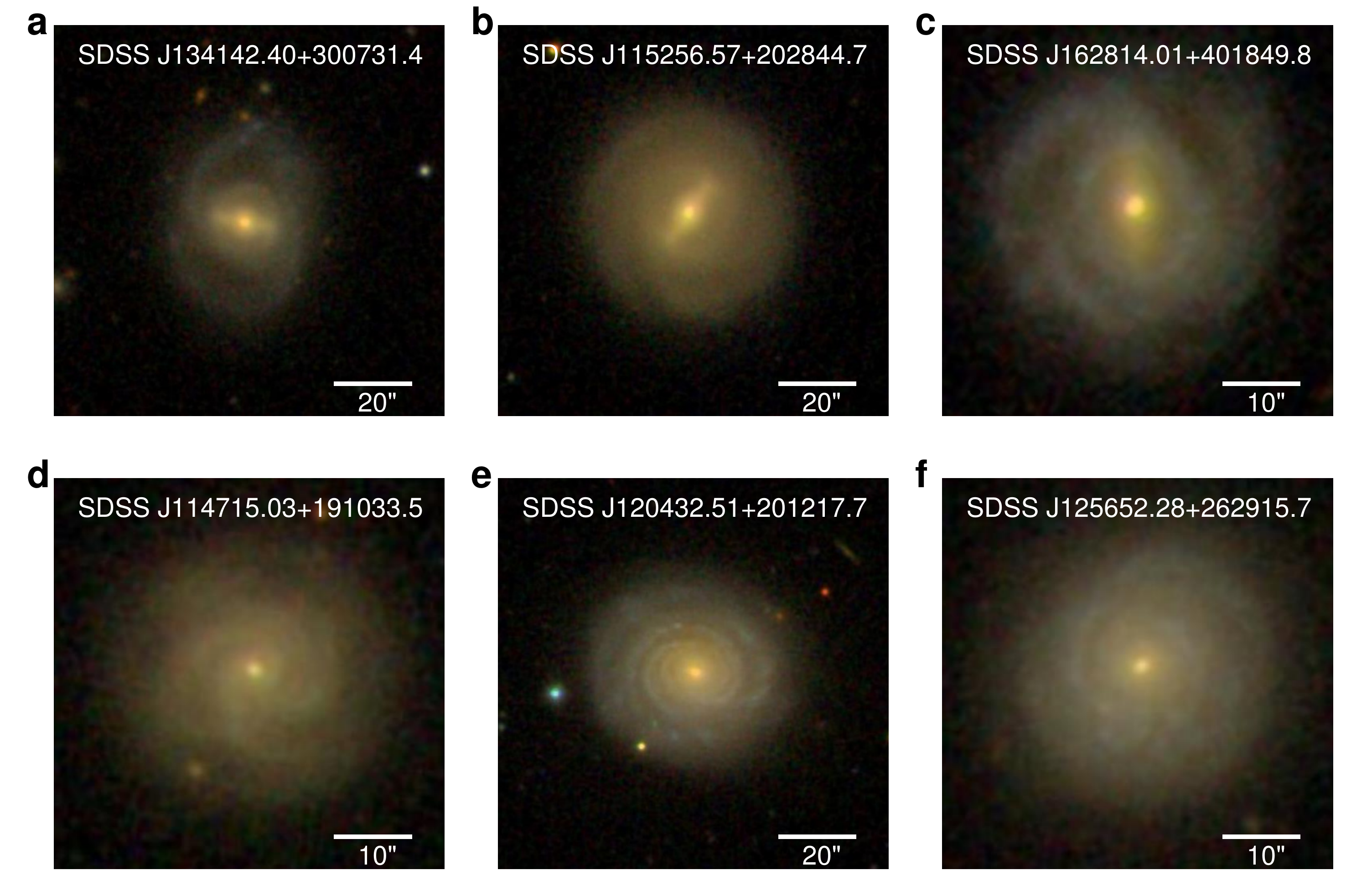}
\centering
\caption{\textbf{Images of barred and non-barred disk galaxies classified in this study.} \textbf{a--c,} Barred galaxies. \textbf{d--f}, Non-barred galaxies. The names of the objects are indicated in each panel. Credit: Sloan Digital Sky Survey (SDSS)
\label{fig:exbar}}
\end{figure*}

  As detailed in Methods, we detected bars in the cluster member galaxies using  a quantitative method that searches for an elongated structure that has a large ellipticity for several consecutive isophotal ellipses but a sudden drop beyond a certain radius, and a nearly constant position angle over only the high-ellipticity region. We complemented the bar classification with a visual inspection to exclude false classifications and add bar galaxies that were not detected by the automated method. This process also examines the radial surface brightness profiles of galaxies, from which we derived the bulge-to-total light ratio ($B/T$) of each galaxy. Example images of barred and non-barred galaxies are presented in Fig. \ref{fig:exbar}.

\begin{figure*}
\includegraphics[scale=0.27,angle=00]{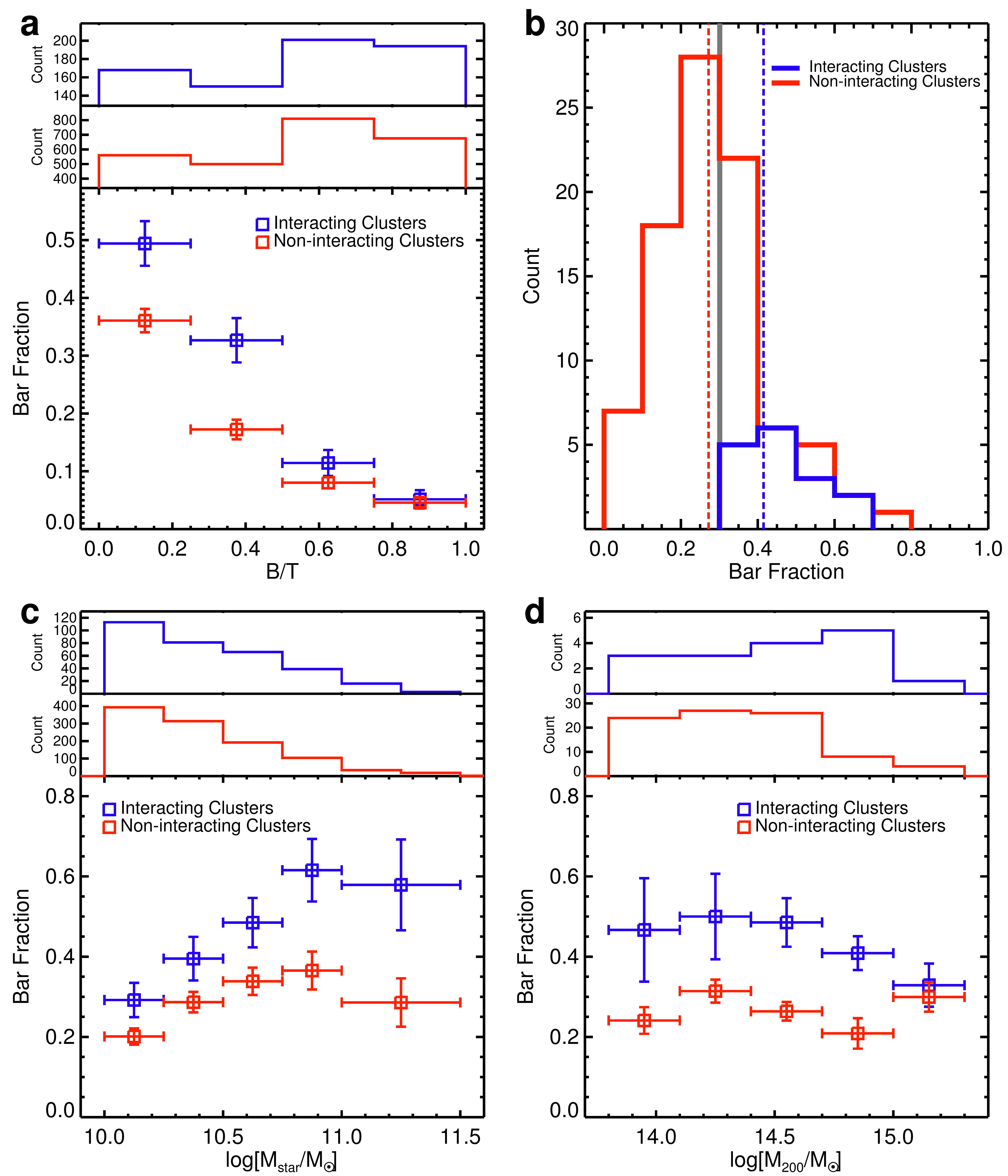}
\centering
\caption{\textbf{Barred galaxies are more abundant in interacting clusters.} \textbf{a}, Bar fractions for interacting clusters and non-interacting clusters as a function of $B/T$ (bottom) with distributions for the number of galaxies contained in each bin (top). The bin size is 0.25, which is indicated with the horizontal error bars. \textbf{b}, Distributions for bar fractions of individual clusters in which only disk-dominated galaxies ($B/T\le0.5$) are used. The total bar fractions for disk-dominated galaxies in the interacting and non-interacting clusters are denoted by the vertical dashed lines. The gray vertical line indicates a bar fraction $f_\mathrm{bar}=0.30$ of late-type galaxies in field (non-cluster) environments (see Methods). \textbf{c}, Bottom, bar fractions of disk-dominated galaxies in the interacting clusters and non-interacting clusters as a function of $M_\mathrm{star}$. Top, $M_\mathrm{star}$ distributions. The bin size is 0.25 dex, and it is indicated by the horizontal error bars. We set the most massive bin to be 0.5 dex due to the lack of the number of such massive galaxies. \textbf{d}, Bottom, the bar fraction of disk-dominated galaxies in the interacting clusters and non-interacting clusters as a function of $M_{200}$ of clusters they belong to. Top, distributions for $M_{200}$. The horizontal error bars of each point indicate the bin size of  0.30. All masses in this figure are normalized to $M_{\odot}$. The error bars of bar fractions (the vertical bars) in all the panels are defined by equation (\ref{eq6}) in the Methods.
\label{fig:bar}}
\end{figure*}

The bar fraction of interacting clusters and non-interacting clusters as a function of $B/T$ is shown in Fig. \ref{fig:bar}a. We find that the bar fraction is several times higher for galaxies with $B/T \le 0.5$ (hereafter, disk-dominated galaxies)  than those with $B/T > 0.5$, in agreement with results from previous studies  \citep{Marinova2009,Barazza2009,Aguerri2009}. The results of previous numerical simulations also support the finding that the formation or maintenance of bars is prohibited in galaxies with high $B/T$ or high central mass concentrations \citep{Shen2004,Athanassoula2005,Bournaud2005}. It is also known that bar formation can be suppressed in galaxies that are dynamically hot (random motions of stars are dominant) or have high velocity dispersions  \citep{Athanassoula1986,Kwak2017}. Given that bars are more commonly found in disk-dominated galaxies, we restrict the analysis of bar fractions to disk-dominated galaxies below. Disk-dominated galaxies are in general late-type or lenticular galaxies \citep{Fukugita1998,Im2002}. There are 1,377 disk-dominated galaxies in our sample clusters.

Interestingly, Fig. \ref{fig:bar}a also shows that the bar fraction is significantly higher in interacting clusters. The significance of the difference is analysed in Fig. \ref{fig:bar}b, where we present the distribution of bar fractions of disk-dominated galaxies in clusters. The vertical dashed lines in the figure show that the total bar fraction of disk-dominated galaxies in interacting clusters is $0.42\pm0.03$ (132/318), whereas the total bar fraction for non-interacting clusters is $0.27\pm0.01$ (288/1059). The total bar fraction is therefore 53\% (or a factor of 1.53) higher in interacting clusters than non-interacting clusters. We tested the significance of the difference of the two distributions using the Kolmogorov-Smirnov test. It is found that the probability ($0\le P \le1$) of the null hypothesis in which the two distributions are drawn from the same distribution is $7.5\times10^{-6}$.  The statistical significance of the difference is very high at $99.999\%$ ($4.5\sigma$). We loosened or tightened the boundary conditions of the criteria for identifying interacting clusters (projected distance difference, radial velocity difference and significance of having substructures) by a factor of 0.6--1.3 to examine how the results change. Even when the criteria are changed, the worst case is that the bar fraction is more than 1.37 times higher in interacting clusters than non-interacting clusters. The statistical significance in such a case is more than $99.9\%$.

 The enhancement of the bar fraction in interacting clusters is ubiquitous over various $M_\mathrm{star}$ from $10^{10}$ to $10^{11.5} \,M_{\odot}$ (Fig. \ref{fig:bar}c) and $M_{200}$ from $10^{13.8}$ to $10^{15.0}\,M_{\odot}$ (Fig. \ref{fig:bar}d).  Over these mass ranges, the bar fraction in interacting clusters is about 1.5 to 2 times higher than that in non-interacting clusters.

This enhancement of bar fraction in interacting clusters compared with non-interacting clusters can be understood through the fact that the rapidly changing time-dependent tidal gravitational fields of interacting clusters impart strong non-axisymmetric perturbations to a disk galaxy and subsequently induce a central stellar bar \citep{Bekki1999}. In principle, both interacting clusters and single clusters exert time-dependent tidal gravitational fields that can induce bar formation, but our finding suggests that interacting clusters offer an environment that is more favourable to bar formation than single clusters. We can explain this behaviour with the following calculation. Assuming that clusters have isothermal sphere mass profiles of $\rho = \rho_{0} r^{-2}$, the rate of change in the tidal forces ($F_\mathrm{tidal}$) exerted on a galaxy that is located at a radius of $R$ from the centre of a single cluster, or the main cluster of two merging clusters, can be written as: 
  \begin{equation}
\frac{dF_\mathrm{tidal}}{dt} \propto \rho_{0} \frac{v_r}{R^3}
\label{eqtidal}
\end{equation}
 where $v_r$ is a velocity component in the radial direction towards the cluster centre (not an orbital speed) and $t$ is time. For a galaxy orbiting in a single cluster ($\sim10^{14} M_{\odot}$) with a semi-major axis of $\sim1$ Mpc and eccentricity of 0.25 or so, the typical $v_r$ would be $\sim100$ km s$^{-1}$. For merging clusters, it is not rare to find an approach speed of two clusters at $\sim1000$ km s$^{-1}$ \citep{Lee2010,Vijayaraghavan2013,Zhang2016}, which can be considered as approximately $v_r$ in this case (orbital motion is neglected). Therefore, at the same $R$, the change in the tidal force is more than 10 times stronger for a galaxy in interacting clusters than in single clusters. As $dF_\mathrm{tidal}/dt \propto v_r/R^3$, the enhancement in the variation of the tidal force by a factor of 10 is equivalent to an increase in the volume ($R^3$) of the tidal force influence for bar formation by a factor of 10. This means that the number of galaxies with tidally induced bars can be 10 times more abundant in interacting clusters than in a single cluster. Furthermore, the rapid change in the tidal force is likely to occur over a more extended area in interacting clusters than in a single cluster. For a single cluster, bars can be induced near the core of the cluster \citep{Lokas2016}, when a galaxy with an orbit with a high eccentricity plunges into the core.  On the other hand, in interacting clusters, this mechanism can form bars not only in galaxies passing through the centres of clusters, but also in galaxies that do not have such highly eccentric orbits or those that orbit the outer part of clusters. 

We investigated another possibility: that the bar fraction enhancement in interacting clusters is due to increased rate of galaxy-galaxy interactions---such as galaxy mergers and fly-bys during cluster interaction \citep{Vijayaraghavan2013}---rather than the tidal-field effect during cluster mergers. This is a valid hypothesis considering that simulations suggest that bars can be induced during galaxy-galaxy interactions \citep{Miwa1998,Berentzen2004,Lang2014}. To test this hypothesis, we examined whether barred disk-dominated galaxies have neighbours that are closer to them in comparison to the other type of galaxies.  To do so,  we calculated the median projected distance from barred and non-barred disk-dominated galaxies in interacting or non-interacting clusters to the nearest galaxy with $\log(M_\mathrm{star}/M_{\odot}) \ge 10.0$ within the rest-frame velocity slice of $\pm3\sigma_{v}$  (here, $\sigma_{v}$ is the velocity dispersion of the cluster that each galaxy belongs to). If the enhanced galaxy-galaxy interaction during cluster interaction is important for inducing bars, we expect that the nearest galaxy distance would be smaller for barred disk-dominated galaxies in interacting clusters than other cases. However, we find that this is not so: the median projected distance to the nearest galaxy to barred disk-dominated galaxies is $143\pm15$ kpc for interacting clusters and $149\pm6$ kpc for non-interacting clusters. The corresponding distances for non-barred galaxies are $110\pm8$ kpc and $128\pm3$ kpc in interacting and non-interacting clusters, respectively. The fact that barred disk-dominated galaxies in both interacting or non-interacting clusters have larger distances to their nearest neighbours than non-barred galaxies indicates that the close galaxy-galaxy interaction may preferentially destroy, not enhance bars. Therefore, we conclude that the bar fraction enhancement in interacting clusters is probably not due to enhanced galaxy-galaxy interactions in such clusters. Moreover, we  note that several previous studies also disfavour bar formation due to galaxy--galaxy interactions on a similar basis \citep{Lee2012,Casteels2013,Lin2014}.

 Besides of the importance of cluster interaction on bar formation, Fig. \ref{fig:bar}c,d shows us other interesting aspects of bar formation. Figure \ref{fig:bar}c shows that the bar fraction increases with $M_\mathrm{star}$ regardless of the cluster status. The increase in bar fraction with $M_\mathrm{star}$ was recognized in previous works \citep{Lee2012,Lee2012b,Masters2012}, and has been suggested to be a result of the abundant gas in low-mass galaxies preventing the growth or maintenance of bars \citep{Masters2012,Davoust2004} either by suppressing dynamical bar instability \citep{Berentzen1998,Villa-Vargas2010} or by transferring angular momentum from the gas to the stellar bar \citep{Bournaud2005}. The increase in the bar fraction enhancement with $M_\mathrm{star}$ (from about 1.5 to 2 times) in interacting clusters can be understood in a similar context, whereby massive galaxies (low gas content) are more easily disturbed to form bars during cluster interactions than low-mass galaxies (high gas content) \citep{Berentzen2004}.

 On the other hand, Fig. \ref{fig:bar}d shows us how plausible it is for a bar to form in a galaxy under the tidal force of a single cluster. According to a theoretical study \citep{Lokas2016}, the tidal force is stronger for more massive clusters and near cluster cores, and thus the bar fraction is expected to increase with the cluster halo mass and concentration. Instead, we see no strong evidence for such a case as a function of $M_{200}$ (Fig. \ref{fig:bar}d) and concentration (see Supplementary Fig. 1). This implies that bar formation through a tidal field within a single cluster alone is not the dominant mechanism inducing bars.

 In addition, Fig. \ref{fig:bar}d also allows us to infer the lifetime of bars. Fig \ref{fig:bar}d shows that clusters with larger masses (which have grown through more mergers) do not have higher bar fraction, although we expect the bar fraction to increase with a number of merging events if the lifetime of the bars is much longer than the merger timescale. Combined with the fact that bar fraction does not increase from field (non-cluster) to cluster environments (Fig. \ref{fig:bar}b), Fig \ref{fig:bar}d suggests that the bar enhancement through cluster interaction is a transient phenomenon that can last for roughly the duration of the cluster interaction. We find that about $15\%$ of clusters are interacting, and this indicates that the lifetimes of the cluster interaction and barred galaxies in such an environment are about 1.5 Gyr, assuming a cluster lifetime of $\sim10$ Gyr. We note that this is shorter than the duration of whole cluster merger process of $\sim5$ Gyr, which has been suggested by some previous studies \citep{Vijayaraghavan2013,Zhang2016,Wetzel2008}. Therefore, there is a possibility that our selection method for interacting clusters is missing clusters at a late stage of merging, and the lifetime of 1.5 Gyr should be considered as a lower limit. Some of the non-interacting clusters have high bar fractions comparable to those of the interacting clusters (Fig. \ref{fig:bar}b), so they might be the clusters at a late stage of merging that were missed by the selection method for interacting clusters.

The fact that barred disk-dominated fraction is $\sim1.5$ times higher in interacting clusters than in non-interacting clusters suggests that the cluster-cluster interaction is plausibly responsible for one-third to one-half of barred disk-dominated galaxies in interacting clusters. Future simulation studies, as well as observational studies with larger sets of samples, should reveal a clearer picture of the enhanced bar fraction in interacting clusters.   
\\

\large
\textbf{Methods}\\
\small
\textbf{Cluster identification and mass measurement.}
We used the MPA-JHU catalogue, which contains all the galaxies with SDSS Data Release 8 (DR8) spectra. This catalogue gives information on galaxy properties such as the stellar mass, redshift, star formation rate, and positions in the sky---many of which are derived from spectral analysis. We focused on the redshift range of $0.010<z<0.065$ where the sample is volume-limited and complete for galaxies with $\log(M_\mathrm{star}/M_{\odot}) > 10.0$. There are 72,185 galaxies that satisfy these redshift and stellar mass conditions. We find that the stellar mass completeness is over $\sim98\%$.

To find cluster candidates from the volume-limited sample of galaxies, we measured the number of galaxies around each galaxy within an aperture of radius 1 Mpc and a rest-frame velocity slice of $\delta v=\pm1000$km s$^{-1}$ centred on the redshift of each galaxy ($N$). We find that a $95.4\%$ percentile of the $N$ is 19, which is also $2\sigma$ ($\sigma=6.9$) above the average $N$ ($5.2$) in the distribution. We regard the galaxies with $N\ge19$ as galaxies in overdense regions. 

 We then linked the galaxies with $N\ge19$ by applying the Friends-of-Friends (FoF) algorithm to combine those galaxies into an overdense region. We used a linking length of 1 Mpc and velocity difference (linking length in the velocity space) of $3000$km s$^{-1}$. For each galaxy (with $N\ge19$) in an overdense region, we measured stellar mass densities within the same aperture and the rest-frame velocity slice as above. Then, we defined the centre of the overdense region (that is, the location of the overdense region) as the location of the galaxy in the highest stellar mass density. Similarly, we set the redshift of the galaxy in the peak of the stellar mass density as the redshift of the overdense region. If an overdense region has multiple galaxies with the same highest stellar mass density, then the galaxy with the highest stellar mass among them is selected for determining the redshift and location of the overdense region. The FoF linking of galaxies with high overdensities can merge two obviously separated overdense regions into one. We examined the overdense regions and found three such cases. Each of them was separated into two regions. Through these processes, we identified 130 overdense regions in the redshift range of $0.015<z<0.060$. This redshift range is smaller than the redshift range of the sample galaxies, to avoid overlooking member galaxies of the overdense regions at the boundary of the redshift range.

For the 130 overdense regions, we measured $M_{200}$ (the cluster mass within $R_{200}$). Here, $R_{200}$ is the radius within which the mean interior density is 200 times the critical density of the universe. First, we selected all galaxies with $\log(M_\mathrm{star}/M_{\odot}) > 10.0$ within a 1 Mpc radius from the centre of the overdense region and applied the FoF algorithm to the rest-frame velocities of these galaxies in the radial velocity space with a linking length (or a velocity gap) of 750 km s$^{-1}$. Then we excluded galaxies that were not connected to the redshift of the overdense region to prevent non-cluster members from being included. Using the remaining galaxies, one-dimensional velocity dispersion, $\sigma_v$, was calculated as the standard deviation of the rest-frame velocity of the galaxies after excluding the galaxies whose rest-frame velocities are beyond $\pm3\sigma$ (3$\sigma$ clipping).

$M_{200}$ and $R_{200}$ were derived from $\sigma_v$ by the method in previous studies \citep{Demarco2010,Kim2016}. $R_{200}$ was obtained by
\begin{equation}
R_{200}=\frac{\sqrt{3}\sigma_v}{10H(z)}
\label{eq1}
\end{equation}
 where $H(z)$ is the Hubble parameter at redshift $z$. $M_{200}$ was derived by
\begin{equation}
M_{200}=3\frac{\sigma_v^{2}R_{200}}{G}
\label{eq2}
\end{equation}

Finally, we defined overdense regions of  $M_{200}>7\times10^{13}\,M_{\odot}$ (or $\log(M_{200}/M_{\odot})>13.85$) as clusters. The total number of clusters is 105. We list the 105 clusters in Supplementary Table 1 with various properties. In this study, galaxies belonging to the cluster (member galaxies) are defined as galaxies within $R_{200}$ from the cluster centre and within a rest-frame velocity slice of $\pm3\sigma_v$ centred on the redshift of the cluster.
 
We examined the reliability of our method of identifying clusters and measuring the cluster masses with the aid of a mock galaxy lightcone catalogue from the GALFORM simulation \citep{Cole2000,Lagos2012}. We only used mock galaxy clusters that have more than or equal to 19 galaxies of $\log(M_\mathrm{star}/M_{\odot}) \ge 10.0$. We restricted the redshift range to $z\le0.15$, to secure sufficient number of mock clusters ($\sim100$ mock clusters) for this test. We also limited masses of the mock clusters to $\log(M_{200}/M_{\odot})>13.85$, where $M_{200}$ of mock clusters is calculated by the equations (\ref{eq1}) and (\ref{eq2}) using all of the galaxies assigned to each cluster (simulation $M_{200}$). The total number of mock clusters is 100. 

 We applied our cluster finding method to the 100 mock clusters. Among them, 91 were detected by the method, giving the detection rate of $91\%$ for our method. For the mock clusters of $\log(M_{200}/M_{\odot})>14.5$, the detection rate was $100\%$. 

 For 91 detected mock clusters, we measured $M_{200}$ by the same method we used to obtain masses of the clusters in the observational data. We then compared the measured $M_{200}$ with simulation $M_{200}$ (see Supplementary Fig. 2). We find that an average difference between the two values is 0.00 dex with a standard deviation of 0.25 dex, barring one outlier beyond $\pm3\sigma$. In conclusion, our method measures $M_{200}$ to an accuracy of 0.25 dex without a significant bias. 
\\

\textbf{Selection of clusters that are in pairs or have substructures.}
 The necessary condition for two clusters of masses $M_{1}$ and $M_{2}$ to be in a bound orbit can be written as:
\begin{equation}
R<\frac{2G\,M_{1}\,M_{2}}{V^{2}\,(M_{1}+M_{2})}
\label{eqbound}
\end{equation}
where $R$ is the physical distance between the clusters, $G$ is the gravitational constant and the velocities ($V$) of the two clusters are assumed to be equal. Assuming that $M_{1}\simeq M_{2} = M$ and $V \simeq \Delta v_\mathrm{pec}$, where the $\Delta v_\mathrm{pec}$  is the absolute peculiar velocity difference of two clusters in the one dimensional space, then equation (\ref{eqbound}) can be written as 
\begin{equation}
R\la10\,\mathrm{Mpc} \times \frac{M/10^{14.0}\,M_{\odot}}{(V/200\,\mathrm{km\,s^{-1}})^{2}}
\label{eqbound2}
\end{equation}
Here, we took $V \approx 200$ km s$^{-1}$ as the plausible value for $V$, considering that $\Delta v_\mathrm{pec}$ averages at about this value from our examination of the GALFORM simulation \citep{Cole2000,Lagos2012} data (average $\Delta v_\mathrm{pec}$ is 193 km s$^{-1}$ with a dispersion of 164 km s$^{-1}$). The mass range of the clusters in our sample is about $10^{14 - 15} M_{\odot}$, meaning that the necessary condition for two clusters in a bound orbit in our sample is $R< 10$ - $100$ Mpc.

 In the rest-frame velocity space, the physical radial distance and the peculiar velocity difference are intermingled. Thus, we use a rough and conservative criterion that two clusters in pairs should have a radial distance between them of less than $\sim10$ Mpc. For the velocity criterion, this translates into  $\Delta v < 750$ km s$^{-1}$, where $\Delta v$ is the absolute difference of the rest-frame velocity of two clusters. We note that $\Delta v_\mathrm{pec}$ is of order of 200 km s$^{-1}$ and the probability of finding two clusters with $\Delta v_\mathrm{pec} > 500 - 600$ km s$^{-1}$ is less than $5\%$ according to our analysis of the simulation data, as well as other observational results \citep{Bahcall1996,Thompson2012}. Therefore, the rest-frame velocity difference of 750 km s$^{-1}$ must reflect the true physical difference of $R\la10$ Mpc for most clusters.

The projected distance criterion for defining clusters in pairs is set to be $D<2\times(R_{200,\,1}+R_{200,\,2}$), where $D$ is the projected distance between two clusters, while $R_{200,\,1}$ and $R_{200,\,2}$ are $R_{200}$ of each cluster. Note that $R_{200}$ is about 0.8 - 2.5 Mpc for cluster samples, so $2\times(R_{200,\,1}+R_{200,\,2})$ has a value in the range of 3.2 - 10 Mpc.  

With these criteria, we identify 7 cluster pairs (hence 14 clusters). Figure \ref{fig:exmap} shows examples of the surface number density maps for clusters in a pair or in isolation. 

 It has been found that interacting clusters along the line of sight are often identified as a single overdensity with substructures in velocity space \citep{Hwang2009,Shim2011,Hou2012}. Therefore, we also identified clusters with substructures and included them in the sample of interacting clusters. To find clusters having substructures, we applied the DS test \citep{DS1988} to the member galaxies of each cluster. This test finds deviations of local mean velocities ($v_\mathrm{local}$) and dispersions ($\sigma_\mathrm{local}$) from velocity of the cluster ($v_c$) and $\sigma_v$. We assigned $\delta_i$ to each member galaxy, which is defined as
\begin{equation}
\delta_i=\frac{(n+1)}{\sigma_v^2}\Big[(v_\mathrm{local}-v_c)^2+(\sigma_\mathrm{local}-\sigma_v)^2\Big]
\label{eq3}
\end{equation}
where $v_\mathrm{local}$ and $\sigma_\mathrm{local}$ were determined from $n$ nearest member galaxies in the projected distance in a cluster. We set $n$ to be the square root of the number of member galaxies in each cluster as in previous studies \citep{Einasto2012,Pinkney1996}. Then, we summed all deviations of the galaxies in a cluster; $\Delta=\Sigma\delta_i$. If a cluster has a substructure (or a subgroup) of galaxies concentrated in a specific area with deviations in the velocity and its dispersion compared with the cluster values, that cluster has a large $\Delta$.

We determined the significance of $\Delta$ by Monte Carlo simulations. We simulated 100,000 trials in which the velocities of galaxies were randomly shuffled, while the velocity distribution of galaxies was maintained and the positions of galaxies were fixed. In the simulations, the random shuffling causes $\Delta$ to be much smaller than the original $\Delta$ if a cluster has true substructures within it. Thus, we can define the significance of having substructures by $P$ value; $P=N(\Delta_\mathrm{sim}>\Delta_\mathrm{obs})/N_\mathrm{sim}$. Here $N(\Delta_\mathrm{sim}>\Delta_\mathrm{obs})$ is the number of simulations where shuffled $\Delta$ ($\Delta_\mathrm{sim}$) is larger than the original value ($\Delta_\mathrm{obs}$), and $N_\mathrm{sim}$ is the total number of trials. The small $P$ value (or the large $1-P$ value) indicates a high probability of existence of substructures in a cluster. 

 We found that five clusters have very high probability of having substructures with the logarithmic $P$ value less than $-4.0$, which means a probability larger than $99.99\%$. We therefore selected these five clusters as clusters having true substructures. Interestingly, three of them are the clusters in pairs. We show examples of the velocity and spatial distributions for the member galaxies of clusters without or with substructures in Fig. \ref{fig:exmap}. 

 In conclusion, we find 16 interacting clusters in total. Of the 16 clusters, 14 clusters are in pairs and 5 clusters have substructures, among which 3 clusters belong to both categories. 
\\

\textbf{Classification of barred galaxy.}
We detected bars in galaxies using a quantitative method that uses the isophote of galaxies. We performed an independent visual inspection to assist the quantitative method. We also obtained surface brightness profiles of galaxies to derive $B/T$ to classify galaxies into those dominated by disks, and determine a quantitative measure of the strength of bulges. 

The IRAF ELLIPSE \citep{Jedrzejewski1987} task was performed on the $r$-band SDSS Atlas images of each galaxy. The Atlas images are the postage-stamp images for individual objects in the SDSS survey. The results of the ELLIPSE task are not sensitive to initial input parameters such as the position angle and the ellipticity once the ellipse fit converges. However, sometimes the ellipse fit does not converge for some initial input parameters. Thus, we fit ellipses to a galaxy multiple times using 36 different sets of input parameters with various position angles and ellipticities. We then abandoned the fitting results that did not converge. We note that the difference of the results depending on the initial parameters is nearly negligible (typical scatter is $\sim0.001$ for ellipticity, while it is $\sim0.3^{\circ}$ for position angle), if the fits converge. 

Ellipses for the outer part of galaxies suffer from low signal-to-noise (S/N) ratios, causing the resultant parameters of the ellipses unreliable. Therefore, we used only the ellipses in outer parts of galaxies whose S/N for intensity is larger than 20 (the error of the surface brightness is less than 0.05) in this study.  The seeing effect is expected for the ellipses in the very central parts of the galaxies, therefore we did not use the ellipses inside a radius of $1\arcsec$ to avoid this effect. We note that the typical value of the half-width at half-maximum (half of the full-width at half-maximum) of the point spread function of the SDSS $r$ band is $0.7\arcsec$.

We failed to fit 18 galaxies among 4,595 galaxies. These 18 galaxies are edge-on galaxies that were not included in the search for barred galaxies. Of the remainder, 105 galaxies were found to be very compact, as the semi-major axes of their outermost ellipses that satisfy S/N$>20$ are less than $4\arcsec$ (less than 2.5 kpc at $z\approx0.03$). We did not include these compact galaxies in detecting barred galaxies, as they are too compact (thus, the small number of data points) to constrain $B/T$ and find bars reliably. Indeed, they were found not to possess bars by the visual inspection.

To derive $B/T$, we fitted the surface brightness profiles of galaxies along the major axis using a model that is a combination of the de Vaucouleurs law for the bulge component and the exponential profile for the disk component. After fitting the model, we obtained the total luminosity of each component using the ellipticity from the ELLIPSE fit and intensity of the model, and then we calculated $B/T$. 

 We compared the $B/T$ values derived by our method to those in the catalogue of \citet{Simard2011}: the average difference between the values is 0.04 (the $B/T$ values from \citet{Simard2011} are slightly larger) and the standard deviation of the differences is 0.21. 

 We only classified bars for galaxies with ellipticities smaller or equal to 0.5 ($e\le0.5$), which is identical to an inclination angle smaller or equal to $60$ ($i\le60^{\circ}$). This is to exclude galaxies with large inclination angles for which it is difficult to determine the existence of bars. Here, we defined the ellipticity of a galaxy as the median ellipticity of the three outermost ellipses with S/N$>20$.  

To detect bars through the ELLIPSE task, we followed a similar method to previous studies \citep{Jogee2004,Marinova2009,Menendez2007}: we identified bars through the radial profiles of ellipticity and position angle. The criteria we used are: (1) $e$ rises above 0.25 for at least three consecutive ellipses (normally at the transition between a bulge and a bar) and reaches a global maximum larger than 0.30; (2) beyond the end of the bar, (or at the bar-to-disk transition) $e$ must drop by more than 0.1 within three consecutive ellipses; (3) the position angle must be maintained within $25^{\circ}$ in the bar region (from the ellipse for which $e$ begins to rise above 0.25 to the ellipse that starts to show a drop of $e$); and (4) in the vicinity of the bar-to-disk transition (where the ellipses show a decreasing trend in $e$ and three additional consecutive ellipses beyond the end of the trend), the position angle changes by more than  $5^{\circ}$. We regarded a galaxy as having a bar only if all of these criteria are met. In Supplementary Figs. 3 and 4, we present examples of barred galaxies with their surface brightness, ellipticity and position angle profiles. 

We complemented the bar classifications yielded from the ELLIPSE task with a visual inspection. This is not only to test the classifications, but also to improve the reliability of the task. We find that $87.7\%$ of galaxies with obvious and prominent bars detected visually were also classified as barred galaxies through the ELLIPSE task. We reclassified the 24 galaxies with obvious and prominent bars that were missed by the ELLIPSE task as barred galaxies. Most of these galaxies missed just one among the four bar classification criteria. 

We also examined the false detection rate of bars by the ELLIPSE task. We find that $13.5\%$ of barred galaxies detected by the ELLIPSE task did not have bars when the images were examined visually. Some of these galaxies have clear dust lanes that mimic bars or are contaminated by foreground sources that disturb the isophotal fit. Other galaxies that were misclassified as barred show a distorted morphology due to close mergers with other galaxies. We reclassified the 82 misclassified galaxies by the ELLIPSE task as non-barred galaxies. We note that results derived from the bar classification without the correction via visual inspection are almost identical to those with the correction. This means that the conclusion of this paper is not greatly affected by the bar classification method.

We compared our bar classifications with those of \citet{Lee2012} (galaxies with axis ratios larger than 0.6). We find that 1,331 galaxies are present in the samples of both studies and among them 89\% (1,181/1,331) of bar classifications coincide with each other.

In this study, the bar fraction is defined as 
\begin{equation}
f_\mathrm{bar}=N_\mathrm{bar}/N_\mathrm{gal}
\label{eq5}
\end{equation}
where $N_\mathrm{gal}$ is the number of galaxies with $e\le0.5$ and $N_\mathrm{bar}$ is the number of barred galaxies among the galaxies with $e\le0.5$. The error for $f_\mathrm{bar}$ is the standard error for the sample proportion for a binomial distribution, which is given as
\begin{equation}
\sqrt{\frac{f_\mathrm{bar}(1-f_\mathrm{bar})}{N_\mathrm{gal}}}
\label{eq6}
\end{equation}

 We calculated $f_\mathrm{bar}$ of interacting clusters and non-interacting clusters based on late-type galaxies of the \citet{Lee2012} catalogue and cluster information ($R_{200}, \sigma_v$) of this study, and compared these values with $f_\mathrm{bar}$ of disk-dominated galaxies ($B/T\le0.5$) of our sample. We note that the $z$ range and sky area coverage of the \citet{Lee2012} catalogue are similar to those of our sample. We find that the barred late-type galaxy fraction of non-interacting clusters based on the \citet{Lee2012} catalogue is $0.32\pm0.03$, whereas that of interacting clusters is $0.43\pm0.04$. Regarding the errors, these fractions agree with those from our sample ($0.27\pm0.01$ and $0.42\pm0.03$ for non-interacting and interacting clusters, respectively).
 
 We also computed $f_\mathrm{bar}$ of the non-cluster environment using the \citet{Lee2012} catalogue. The non-cluster environment was defined as all regions outside of $R_{200}$ from the cluster centres and rest-frame velocity slices of $\pm3\sigma_v$ centred on the redshifts of the clusters. The $f_\mathrm{bar}$ value of the non-cluster environment is 0.30, which is almost the same as that of non-interacting clusters.
\\

\normalsize
\textbf{Data availability}\\
\small
The data used in this work can be downloaded from the public data archive of SDSS (\url{http://skyserver.sdss.org}) and MPA-JHU catalog at the website of \url{https://www.sdss.org/dr14/spectro/galaxy_mpajhu/}. The other data used for the plots within this paper are available from the corresponding author upon reasonable request.
\\

\normalsize
\textbf{Acknowledgements}\\
\small
We thank Woong-Tae Kim and SungWon Kwak for helpful comments and discussions.
This work was supported by the National Research Foundation of Korea (NRF) grant, No.
2017R1A3A3001362, funded by the Korea government (MSIP). G.-H.L. was supported by KASI-Arizona Joint Postdoctoral Fellowship Program jointly managed by Korea Astronomy and Space Science Institute (KASI) and the Steward Observatory, at the University of Arizona.
\\

\normalsize
\textbf{Author contributions}\\
\small
Y.Y. led the analysis. Y.Y. and M.I. led the interpretation and wrote the paper. G.-H.L. and G.L. contributed to the classification of bar galaxies. G.-H.L. provided scientific inputs on the issues related to the morphology classification. S.-K.L. contributed to the cluster search. All authors contributed to the discussion of the results, reviewed the paper, and provided input on the manuscript.
\\

\normalsize
\textbf{Competing interests}\\
\small
The authors declare no competing interests.
\\

\end{document}